# A self-similar morphology detected in a composite produced by densification of co-crumpled metallic thin foils

Olivier Bouaziz[1,2,*], Yuri Estrin[3,4], Yan Beygelzimer[5], Roman Kulagin[6]

[1]Laboratoire d'Etude des Microstructures et de Mécanique des Matériaux (LEM3), CNRS, Université de Lorraine, Arts et Métier Paris Tech, F 57000, Metz, France
[2]LABoratoire d'EXcellence DAMAS, Université de Lorraine, 57000, Metz, France
[3]Department of Materials Science and Engineering, Monash University, Clayton, Australia
[4]Department of Mechanical Engineering, The University of Western Australia, Perth, Australia
[5]Donetsk Institute for Physics and Engineering named after A.A. Galkin, National Academy of Sciences of Ukraine, Kyiv, Ukraine
[6]Institute of Nanotechnology, Karlsruhe Institute of Technology, Eggenstein-Leopoldshafen, Germany
* Email: olivier.bouaziz@univ-lorraine.fr

Abstract

A new kind of composite were manufactured by densification of co-crumpled aluminium and tantalum thin foils using close die compression. It was shown by optical micrography that its microstructure is highly interlocked. The morphology was analysed quantitatively in terms of the following three parameters: the area of the foil interface per unit volume, the interface tortuosity, and a characteristic of the local orientation of the foil surface. Based on these parameters, co-crumpled material studied has been compared with conventional laminates. A fractal nature of its self-similar structure was revealed.

1. Introduction

One of the most promising approaches to design of new materials is architecturing of the inner geometry of a material at a length scale intermediate between the microstructural and the macroscopic ones [1]. This extends the possibilities to manipulate the properties of materials by using their meso scale architecture as a new design parameter [2]. Of special interest are materials whose desired architecture is achieved by self-organisation, without specific external actions [3, 4]. In this article we present a new class of material architectures, which is produced by co-crumpling of thin foils of different metals followed by compaction in a closed die. Their mesostructure, exemplified by co-crumpled aluminium and tantalum foils, is analysed based on the parameters proposed to characterise the geometrical features of the inner architecture of the material.

2. Manufacture

As the initial materials, two A4 format thin foils of aluminium (thickness of 18µm) and tantalum (thickness of 10µm), both of commercial purity, were used. The two foils were co-crumpled manually. The tangle thus obtained was densified by close-die compression. The maximum value of compression force for obtaining a sample was 30 tonnes. The samples produced, 10 mm on diameter, exhibited a strong cohesion of the constituents. The microstructure was characterised by optical microscopy at different orientations, as shown in Fig. 1.

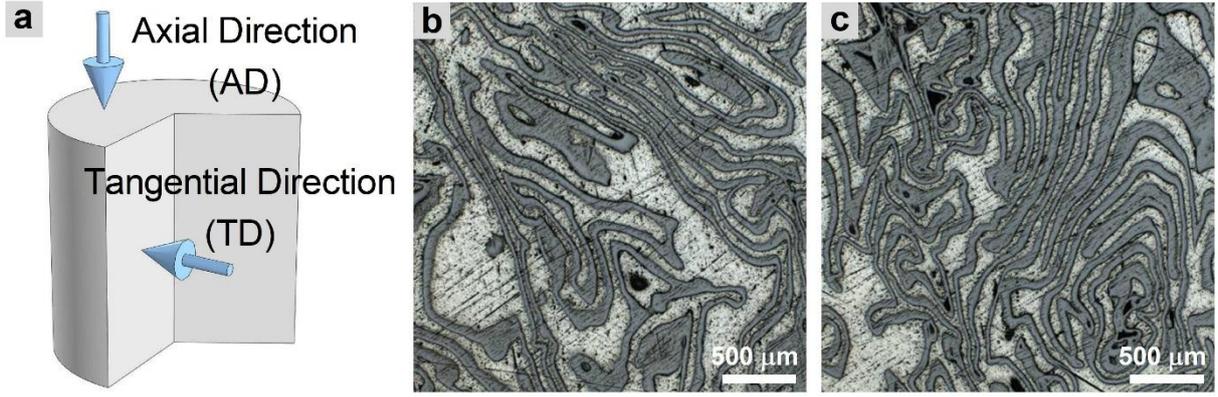

Fig1. Optical microscopy after densification for different orientations (aluminium: light, tantalum: dark): (a) schematics of the sample, (b) structure observed in a cross-section normal to AD, (c) structure in a cross-section normal to TD.

3. **Quantitative microstructure analysis**

The architecture of a compacted crumpled material (CCM) is unquestionably stochastic, cf. Fig 1b,c. Rigorous methods for characterising such patterns are being developed in stochastic geometry [5]. Below we propose three parameters of the architecture of CCMs, which in our opinion are the key characteristics determining their mechanical properties.

The first parameter describes how developed the network of the interfaces of a CCM is. This characteristic is captured by the parameter $\bar{S}$ well known in the stereology. It is defined as the total area of the surface of a phase per unit volume of the material [6]. As the area $S$ of an initial foil remains unchanged when it is crumpled, one has

$$\bar{S} = \frac{S}{V}, \qquad (1)$$

where $V$ is the sample volume. For a fully dense sample, this volume is given by the total volume of the co-crumpled foils:

$$V = Sh, \qquad (2)$$

where $h$ denotes the sum of the foil thicknesses.

Combining Eqs. (1) and (2) yields

$$\bar{S} = h^{-1}, \tag{3}$$

It is interesting to compare a CCM with a laminate material (LM) comprised by alternating Al and Ta foils with respect to this parameter. For an LM, the area of interfaces, $s$, within the sample volume can be estimated as $s = wn$, where $w$ is the area of a horizontal cross-section of the sample and $n = H/h$. As the sample volume is given by $V = wH$, it follows from Eq. (1) that the specific interface area, $\bar{S}_{LM}$, for an LM is expressed by $\bar{S}_{LM} = h^{-1}$, and is thus identical with the corresponding quantity for a CCM, cf. Eq. (3). That is to say, in terms of this characteristic, a CCM is not different from an LM.

The second parameter represents the tortuosity of the structure, which strikes the eye in Fig. 1b,c. In geometry, tortuosity of a curve is defined as the ratio of its length between two points to the length of a straight line connecting them [7]. Following this definition, the mean value of the tortuosity $T$ of a CCM can be represented by

$$T = \frac{D}{d}, \tag{4}$$

where $D$ and $d$ are, respectively, the characteristic dimensions of the blank (the foil) and the CCM sample.

This parameter can be expressed in a form that contains characteristics of the final sample only. Considering that the sample volume is equal to the sum of the volumes of the foils and using for them order-of-magnitude estimates, $d^3$ and $D^2h$, respectively, one obtains from Eq. (4):

$$T \sim \sqrt{\frac{d}{h}} \tag{5}$$

Setting, as an example, the values of $h = 10\ \mu m$ and $d = 10\ mm$, one obtains from Eq. (5) the following estimate for a CCM: $T \sim 30$. Since the interfaces in an LM are planar, their tortuosity is $T_{LM} = 1$.

It is known that the interfaces between the layers can prevent the propagation of cracks in a composite [8]. This gives reasons to believe that a CCM, with its much more developed tortuosity of foil interfaces, will possess better strength characteristics than an LM.

The third parameter of the architecture, which we believe affects the mechanical properties of a CCM, characterises the local orientation of the foil surface. The broader is the foil segment located within an axial cross-section of the foil, the smaller is its inclination angle with the axis. This is reflected in the non-uniformity of the distribution of the dark and light regions in the microstructures seen in Fig. 1. The non-uniformity of the structure can be quantified by using the mixing index $Q$ proposed in Ref. [9]. As applied to a two-component mixture, this index can be calculated in the following way. An area of interest with a characteristic size $L$ is subdivided into $N$ elements of size $l < L$. In each element $k$, the concentration of one of the components, $C_k (k = 1,2, ..., N)$ is determined. The index $Q$ is then calculated according to the formula

$$Q = \sqrt{\frac{\Delta G}{\Delta G_{max}}}, \qquad (6)$$

where $G = \frac{1}{N}\sum_{k=1}^{N} C_k^2$ is the Gibbs parameter, $\Delta G = G - C^2$, $\Delta G_{max} = C - C^2$, and $C$ is the concentration of the component considered in the entire area of interest.

The parameter $Q$ characterises the uniformity (or non-uniformity) of the composition over elements of size $l$: when the concentration of the components in each element is equal to that in the entire body, $Q = 0$ holds. In the opposite limit case when the components of a mix are fully separated, i.e. $C$ is equal to either zero or unity, $Q = 1$ follows. In the context of CCMs, the quantity $Q$ represents the degree of uniformity of the orientation of the foil surface within the bulk of a sample.

The magnitude of $Q$ depends on the element size $l$. Fig. 2 displays this dependence for the structure shown in Fig.1b,c and an idealized multilayer structure composed of appropriately sized aluminium and tantalum foils.

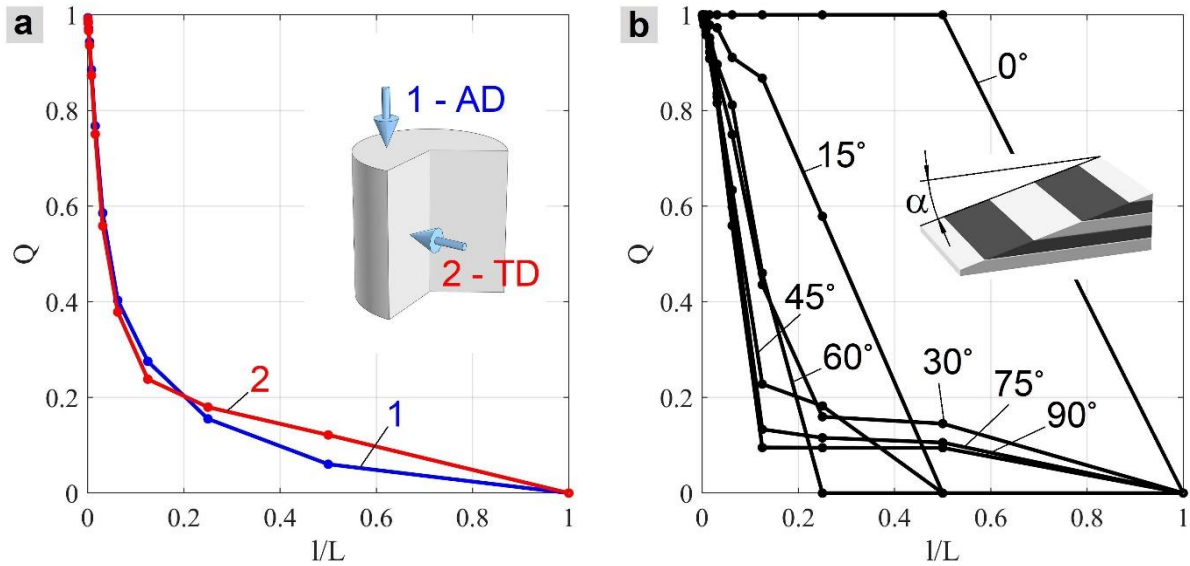

Fig. 2. The mixing index $Q$ as a function of the ratio $\left(l/L\right)$ for the structures considered: (a) CCM depicted in Fig.1b,c; (b) idealized layered structure. The inclination angle of the cross-section chosen for viewing is shown for clarity.

Figure 2 shows that the index $Q$ of a CCM is practically independent of the orientation of the sample cross-section chosen for viewing, whereas for an LM it can be radically different for different orientation angles of the cross-section. This reflects anisotropy of LMs and indicates the largely isotropic character of CCMs.

It is known from literature that crumpled structures display fractality [10]. Self-similarity, or fractality, of the structure is further supported by the value of the corresponding Hausdorff-Besicovitch dimension [11]. It is found from the slope of the straight line representing the dependence of the number of elements containing dark fragments of the structure on the element size in double-logarithmic coordinates. Figure 3 displays this for the present CCM.

The graphs indicate clearly that the CCM investigated possesses a fractal structure. The fractal dimension given by the Hausdorff-Besicovitch dimension, $\dim_H$, is shown in Fig. 3 for both viewing cross-sections studied. The close values of $\dim_H$ found for both cross-section orientations is a further substantiation of the material's isotropy.

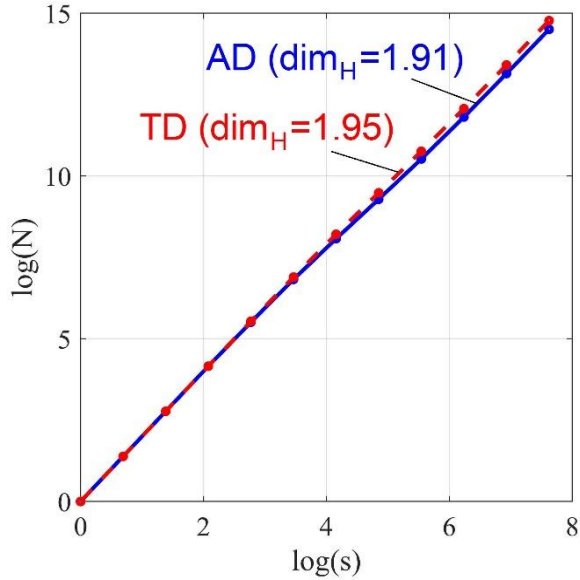

Fig. 3 Dependence of the number of elements containing dark fragments of the structure presented in Fig. 1b,c on the element size in double-logarithmic coordinates. The two data sets correspond to the observation planes normal to the sample axis and to the transverse direction. The Hausdorff-Besicovitch dimension is shown for both cases.

### 4. Conclusion

In conclusion, we would like to point out the following. The functional dependence of the mechanical properties of CCMs on the parameters or their architecture is to be determined from experiment. From general considerations one can conjecture that with increasing $\bar{S}$ and $T$ the yield strength, the ultimate tensile strength, and the fracture toughness of CCMs will increase. The above analysis has shown that the third characteristic parameter, the mixing index $Q$, provides an insight into subtle features of the CCM architecture. The effect of this index on the mechanical properties of CCMs is yet to be explored.

Acknowlegment

The authors thank B. Bouaziz for prompting this study. YE, YB, and RK acknowledge support from The Volkswagen Foundation through a Cooperation Project (Az.:97 751).